\documentstyle[prl,aps,hcm,epsf,floats]{revtex}
\epsfclipon
\begin{document}
\draft
\title{
Spontaneous Interlayer Charge Transfer near the Magnetic Quantum Limit
}
\author{H. C. Manoharan,$^{1,*}$ Y. W. Suen,$^1$ T. S. Lay,$^1$ 
        M. B. Santos,$^1$ and M. Shayegan$^{1,2}$}
\address{$^1$Department of Electrical Engineering, Princeton University,
    Princeton,  New Jersey  08544, USA\\
    $^2$Sektion Physik, LMU M\"unchen, Geschwister-Scholl-Platz 1, 80539 
    M\"unchen, Germany
}
\date{Phys. Rev. Lett. {\bf 79}, 2722 [1997]; Submitted 13 June 1997; Accepted 14 August 1997}
\maketitle

\vspace{-0.18cm}

\begin{abstract}

Experiments reveal that a confined electron system with
two equally-populated layers
at zero magnetic field can spontaneously break this symmetry through an 
interlayer charge transfer near the 
magnetic quantum limit.  New fractional quantum Hall states at unusual
total filling factors such as $\nu = \frac{11}{15}$ 
$\left(=\frac13 + \frac25\right)$ stabilize as signatures that the system
deforms itself, at substantial electrostatic energy cost, in order to
gain crucial correlation energy by  
``locking in'' separate incompressible liquid phases
at unequal fillings in the
two layers (e.g., layered $\frac13$ and $\frac25$ states in the 
case of $\nu = \frac{11}{15}$). 

\end{abstract}


\twocolumn

Imagine creating a high-quality two-dimensional (2D) electron system
and then moving in from afar a second identical layer constrained to a
parallel plane.  As the separation between the two charged sheets shrinks, 
both Coulomb repulsion and electron tunneling between the layers 
increase.  Now imagine, after establishing suitably close proximity between
the two 2D systems, turning a knob that quenches the interlayer electron 
tunneling.  What remains is a symmetric bilayer system in which 
charge transfer between layers, in analogy to a parallel-plate capacitor, 
can only occur by surmounting a sizable electrostatic energy barrier---the 
``out-of-plane'' Hartree charging energy.  For a system of sufficiently
low density, this charging energy may be overcome by the exchange-correlation
energy, leading to a spontaneous transfer of all charge into one of the 
layers.  Such so-called ``exchange-instabilities'' have been predicted
to occur in multilayers
%\cite{MacDonald:PRB:Staging,Ruden:APL:Bilayer}
\cite{Instabilities}
but have not yet been observed experimentally
\cite{Papadakis:PRB:Bilayer}.  This may be due to 
the fact that samples of sufficiently low density and low disorder
have not been available, or possibly because other more exotic bilayer
phases \cite{Zheng:PRB:Exchange}
intervene and maintain a lower energy than the 
spontaneously-generated monolayer phase.  

Envision, on the other hand, approaching the diametric 
regime---the extreme magnetic quantum limit.  It is known
that when a large magnetic field $B_\perp$ is applied
perpendicular to the plane of a 2D electron layer, electron-electron
interactions can dominate and lead to new ground states such as the
fractional quantum Hall (FQH) liquids %\cite{Tsui:PRL:FQHE} 
at certain Landau-level filling factors $\nu$.  Because of the
particular stability (cusp-like minima in energy
\cite{Halperin:PRL:FQHE}) of the FQH phases at these special fillings,
one might expect a partial charge transfer between the two layers of
an interacting bilayer electron system (ES) if the ensuing
inequivalent layers can each support a strong FQH state.  Here we
report strong experimental evidence that such a charge transfer indeed
occurs.

By modulation-doping a 750 \AA -wide GaAs quantum well, we fabricated
a special ES whose density $n$ and charge distribution are controlled
via a pair of metal front- and back-side gates
\cite{Suen:PRB:WQW,Suen:PRL:Origin,Manoharan:PRL:BiWC}.  Measurements
were per-\break\vspace{5.9cm}

\noindent
formed at base temperatures $T \simeq 30$ mK.  For symmetric
(``balanced'') charge distributions, the ES in a wide well can, in
general, be tuned from a single-layer to an interacting bilayer system
by either increasing $n$ or applying an in-plane magnetic field
$B_\|$.  Increasing $n$ results in an increase (decrease) of the
charge density near the sides (center) of the well, and reduces the
interlayer tunneling.  For a fixed $n$, increasing $B_\|$ (by
increasing the angle $\theta$ between the normal to the sample plane
and the magnetic field direction) has a qualitatively similar effect
\cite{Hu:PRB:Bpara,Lay:Tilt2/3}.  The evolution of the FQH states in
this system with increasing $n$ has been reported recently
\cite{Suen:PRL:Origin,Manoharan:PRL:BiWC}.  At the lowest $n$ the
sample exhibits the usual FQH effect at exclusively odd-denominator
$\nu$, while at the highest $n$ the strongest FQH states are those
with {\it even numerators}, as expected for a system of two 2D ESs in
parallel.  For intermediate $n$, {\it even-denominator} FQH states at
total fillings $\nu = \frac12$ and $\frac32$, which are stabilized by
both interlayer and intralayer correlations, are observed.  Also of
relevance to this paper is the observed one-component (1C) to
two-component (2C) transition of certain FQH states such as those at
$\nu = \frac23$ and $\frac43$.  Note that the $\frac23$ state, for
example, can exist as a FQH ground state of the entire electron layer
in the wide well (1C), or as two parallel $\frac13$ FQH states in the
separate layers (2C) \cite{Explain1C2C}.  Indeed the $\frac23$ and
$\frac43$ FQH states in this ES exhibit a clear 1C to 2C phase
transition with increasing $n$
\cite{Suen:PRL:Origin,Manoharan:PRL:BiWC} or $\theta$
\cite{Lay:Tilt2/3}, as evidenced by a pronounced minimum in each of
their quasiparticle excitation gaps measured as the system is tuned
from a monolayer to a bilayer.

Of central interest here is the appearance of FQH states at unusual
$\nu$ in this ES.  We have observed such states near the $\nu =
\frac23$ and $\frac43$ FQH states once these become of 2C origin.  The
data of Fig.\ 1,
\begin{figure}[t]
%\vspace{-1.4cm}
\epsfxsize=\hsize \epsfbox{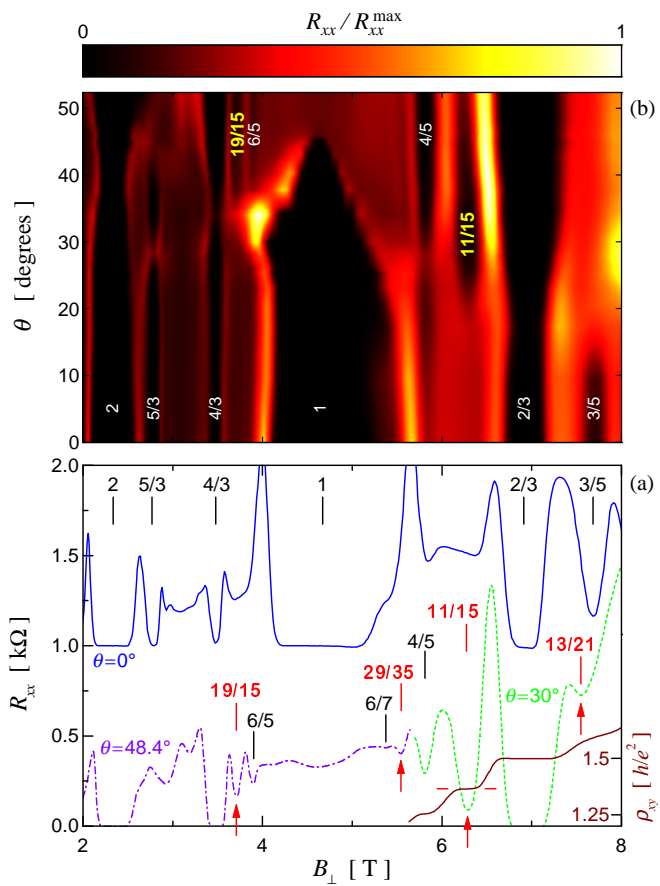}
\vspace{0.15cm}
\caption{Tuning the bilayer ES (at fixed 
  $n = 11.2 \times 10^{10}$ cm$^{-2}$) from the 1C to the 2C regime by
  increasing $\theta$ (hence $B_\|$).  (a) $R_{xx}$ vs $B_\perp$
  ``slices'' through the image of (b), where the normalized $R_{xx}$
  is mapped to a grayscale and also plotted vs $\theta$ (total
  fillings $\nu$ are labeled).  Spontaneous interlayer charge transfer
  engenders new FQH liquids (see features marked by vertical arrows
  and bold fractions).  }
\label{fig1}
\vspace{-0.42cm}
\end{figure}
taken at fixed $n = 11.2 \times 10^{10}$ cm$^{-2}$
with variable $\theta$, provide an example.  The top curve in Fig.\ 
1(a) shows that at $\theta = 0$, the ES exhibits FQH states which are
primarily single-layer-like.  However, upon tilting the sample to
$\theta = 30^\circ$ and $48.4^\circ$ (bottom $R_{xx}$ traces), several
additional FQH states appear between $\nu = \frac23$ and $\frac43$
while the integral QH effect at $\nu = 1$ disappears.  Considering
that at these large $\theta$ values the ES (for $\nu \leq 2$) has
become bilayer-like, some of the additional FQH states---such as those
at $\nu = \frac45$, $\frac67$, and $\frac65$---can be understood as
the descendants of the $\nu = \frac23$ and $\frac43$ states.  For
example, assuming that the 2C $\nu = \frac23$ state arises from two
$\frac13$ FQH states in each layer, the FQH sequence at $\nu =
\frac23$, $\frac45$, $\frac67$ is simply the usual $\frac13$,
$\frac25$, $\frac37$ sequence for each layer.  The well-developed FQH
states at $\nu = \frac{11}{15}$ and $\frac{19}{15}$ (see vertical
arrows), however, cannot be accounted for in any reasonable bilayer or
single-layer scheme \cite{Monolayer}.  On the other hand, note that
each of these unusual fractions can be expressed as the sum of two
simpler fractions.  In fact, this is not a mere coincidence, and our
data indicate that at these special $\nu$, the system is unstable
toward an interaction-induced interlayer charge transfer that ``locks
in'' independent FQH states at unequal fillings in each layer.  At
$\nu =\frac{11}{15}$, for example, one layer has $\frac13$ filling and
the other $\frac25$.  Similarly, the FQH state at $\nu =
\frac{19}{15}$ $\left(=\frac23 + \frac35\right)$ can arise from an
interlayer charge transfer leading to FQH states at $\frac23$ and
$\frac35$ fillings in two layers.  In the remainder of the paper we
elucidate the behavior of these unusual states as a function of
$\theta$ (at fixed $n$), $n$ (at fixed $\theta$), and intentionally
imposed charge transfer (at fixed $n$ and $\theta = 0$).

Figure 1(b) summarizes the evolution of the FQH states in this ES as a
function of increasing $\theta$.  We have condensed a large set of traces
onto the ($B_\perp,\theta$) plane by mapping $R_{xx}$ (normalized
to its maximum value within the plotted parameter range) to a grayscale 
color between black and white.  In such a plot, the QH and FQH phases 
show up as dark black regions, whose width along the $B_\perp$ axis
is a reflection of the strength of the associated state, 
i.e.\ the magnitude of its energy gap.  
The traces in Fig.\ 1(a) can be interpreted 
as constant-$\theta$ slices through the image of Fig.\ 1(b).  As 
$\theta$ is increased, the system is swept from the 1C
through the 2C regime;  a visible measure of this general evolution is 
the weakening and eventual collapse of the $\nu = 1$ QH state.  
The $\nu = \frac35$ FQH state, another 1C state,
is also destroyed by the increasing $B_\|$.
For the states that can exist as both 1C and 2C phases, transitions between
the two ground states are evident.  For example, the $\nu = \frac23$
and $\frac45$ states undergo a 1C to 2C transition at
$\theta \simeq 18^\circ$ and $27^\circ$, respectively.  
Nestled between these two states and in close proximity 
to their 1C$\leftrightarrow$2C transitions,
an $\frac{11}{15}$ FQH state develops and becomes quite strong.
At the same time, $\rho_{xy}$ exhibits a quantized plateau at
$\frac{15}{11} (h/e^2)$ [see lower right of Fig.\ 1(a)].
Very similar behavior is observed on the other side of $\nu = 1$ 
in Fig.\ 1(b).
A $\frac{19}{15}$ state develops in the vicinity of the $\nu = \frac43$
1C$\leftrightarrow$2C transition (at $\theta \simeq 35^\circ$) along with
the appearance of the 
2C $\frac65$ state (at $\theta \simeq 38^\circ$). 

\vspace{-0.016cm}

We have observed a similar evolution of the $\nu =
\frac{11}{15}$ state in the same sample at $\theta = 0$ as a
function of increasing $n$.  While at low $n$ the FQH states
have single-layer origin, when the system is tuned to
very high $n$, as shown in Fig.\ 2, the dominant states are of
2C origin, as evidenced by the preponderance of even-numerator
FQH states.  Of special note, however, are the states marked with
vertical arrows that do not fit in the normal 1C or 2C hierarchy.
Like the 2C case at high $\theta$ described above, states here at
$\nu = \frac{11}{15}$ and $\frac{19}{15}$ are very strong and are
accompanied by Hall plateaus as indicated.   
For $\theta = 0$, the $\nu = \frac{11}{15}$ FQH state starts to
develop at $n \simeq 13 \times 10^{10}$ cm$^{-2}$, slightly
above the $n$ at which the $\nu = \frac23$ FQH state makes
its 1C to 2C transition \cite{Suen:PRL:Origin,Manoharan:PRL:BiWC}.

\begin{figure}
\vspace{-0.8cm}
\hglue -0.6cm
\epsfxsize=9.2cm \epsfbox{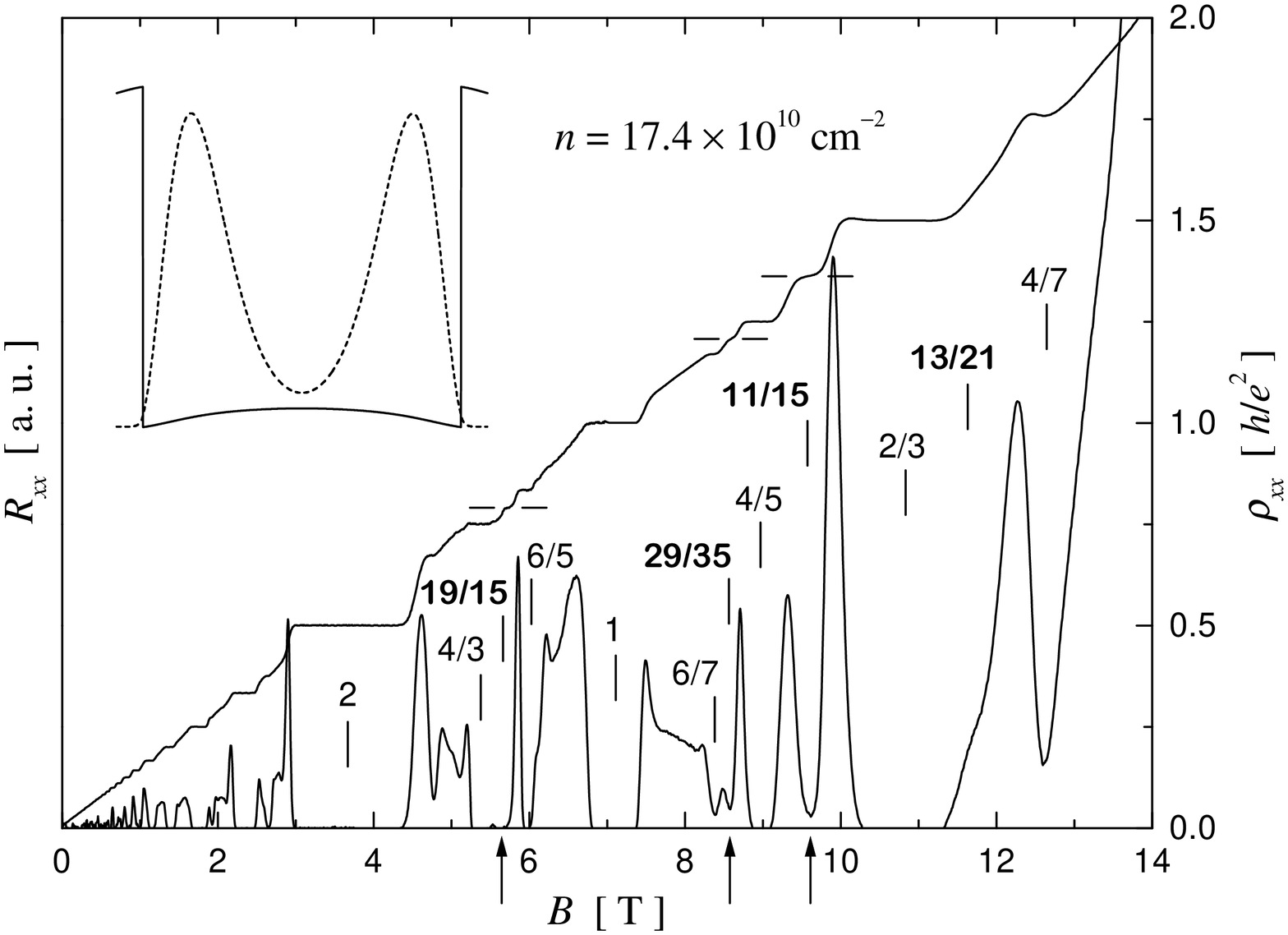}
\vspace{-0.37cm}
\caption{Bilayer ES at $\theta = 0$, tuned deep into the 2C regime by
biasing to high $n$. Interlayer charge transfer becomes favorable
under these conditions as well (see marked FQH states).
Inset: Zero-$B$ charge distribution at this density.}
\label{fig2}
\vspace{-0.37cm}
\end{figure}

\vspace{-0.02cm}

For conciseness, 
let us concentrate on the strongest of these ``special'' 
states, namely the one at $\nu = \frac{11}{15}$. 
To summarize the data presented so far, we observe a FQH state at
$\nu = \frac{11}{15}$ in an ES in a wide quantum well under
either of two conditions:  (a) without an in-plane $B$ but at
large $n$ so that the ES at and near $\nu = \frac23$ has
already made a 1C to 2C transition; (b) starting with a 1C state (i.e.,
low enough $n$) but applying a sufficiently large  
$B_\|$ so that again the ES at and near $\nu = \frac23$ has made a
1C to 2C transition.  We emphasize that in both cases, in the absence
of any applied $B$, the ES has a symmetric (``balanced'') charge
distribution.  Since in both cases the $\frac{11}{15}$ FQH state
appears when the system has become 2C, it is reasonable to assume that
tunneling is sufficiently reduced and can be ignored \cite{Tunneling}.

Thus, ignoring tunneling, we concentrate on the competition between two
energies: 

\noindent (1) {\sl Energy cost} ---
The capacitive energy ${\cal E}_{\text{CAP}}$
which works against the formation
of a $\left( \frac13 + \frac25 \right)$ 
state as it opposes interlayer charge
transfer.  A rough estimate of this energy, expressed per electron, 
for the system at $n = 11.2 \times 10^{10}$ cm$^{-2}$ is  
\begin{equation}
\label{Ecap}
{\cal E}_{\text{CAP}} \simeq {Q^2 \over 2 C}\cdot {1\over n} =
    {(e n_t)^2 \over 2 (\epsilon / d) n} \lesssim 1\, \text{K},
\end{equation} 
where  $Q \equiv e n_t = 5.1 \times 10^9$ $e/\text{cm}^2$
is the charge transfer per unit area from one layer to the other
necessary to set up the
$\left(\frac13 : \frac25 \right) = (5:6)$ layer density ratio,
$C \equiv \epsilon / d$ is the interlayer
capacitance (per unit area), $d \simeq 480$ \AA\ is the interlayer
spacing, and $\epsilon = 13.1$ is the GaAs dielectric constant.
Equivalently,
${\cal E}_{\text{CAP}}$ is the electrostatic energy difference between
the balanced and imbalanced charge configurations pictured in the
insets to Fig.\ 3.

\begin{figure}
\vspace{-0.99cm}
\centerline{
\epsfxsize=9cm \epsfbox{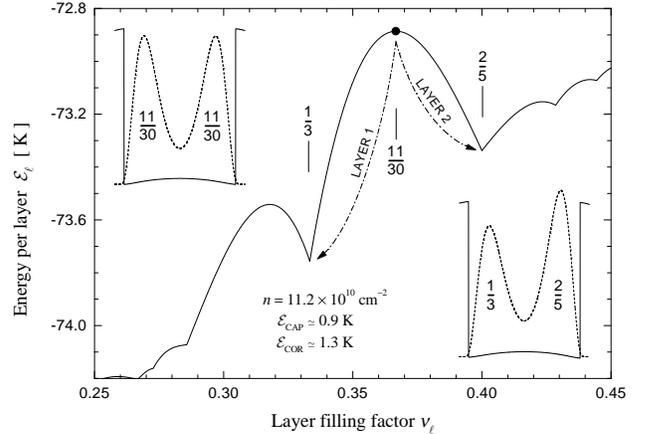}
}
\vspace{-0.025cm}
\caption{Competition between capacitive charging energy 
$\cal E_{\text{CAP}}$ (insets) and
liquid correlation energy $\cal E_{\text{COR}}$ 
(arrows) governing the susceptibility
toward spontaneous interlayer charge transfer.  Displayed
fractions are layer fillings $\nu_\ell$.}
\label{fig3}
\vspace{-0.34cm}
\end{figure}

\noindent (2)  {\sl Energy savings} ---
The correlation energy ${\cal E}_{\text{COR}}$
gained by forming the incompressible
${\frac13}$ and ${\frac25}$ FQH liquid states in the two separate
layers.  To estimate this energy, we have calculated the total energy
of a single layer (i.e.\ half of the bilayer system) 
as a function of the layer filling factor $\nu_\ell$.  The results are
plotted in Fig.\ 3 and are based on the 
best numerical calculations available to date \cite{Price:PRB:XC} for the
exchange-correlation energies and quasiparticle excitation gaps of 
principal FQH states in a realistic (i.e.\ of finite
density and non-zero thickness) 2D system.
Higher-order FQH states were added to the calculation by exploiting   
the energy gap scaling law \cite{HLR,Manoharan:PRL:CF}
$\Delta _\nu \propto (e^2/\epsilon \ell _B)/(2mp+1)$, where
$\Delta_\nu$ is the quasiparticle excitation gap of 
the $\nu = p/(2mp+1)$ FQH liquid ($p$ integer, $m=1,2,\ldots$), and 
$\ell_B \equiv \sqrt{\hbar /eB_\perp}$ is the magnetic length.
Appropriate corrections for Landau level mixing and 
finite thickness were assumed:
for a bilayer system of total $n = 11.2 \times 10^{10}$ cm$^{-2}$,
a layer of density $n/2$ has $r_s \simeq 2.3$ (average interparticle
spacing normalized to the effective Bohr radius) and half-width
$\lambda \simeq 85$ \AA.  The spline procedure described in 
Ref.\ \cite{Price:PRB:XC} was used to connect the various states, rendering
the cusps evident in the energy per layer 
${\cal E}_\ell\left(\nu_\ell\right)$
shown in Fig.\ 3.  The liquid correlation energy gain is depicted by
the peak-to-trough arrows in Fig.\ 3; its magnitude (again, per electron) is
\begin{equation}
\label{Ecor}
{\cal E}_{\text{COR}} \simeq 
    2 {\cal E}_\ell|_{\nu_\ell=\frac12 \cdot \frac{11}{15}} - 
    {\cal E}_\ell|_{\nu_\ell=\frac13} - {\cal E}_\ell|_{\nu_\ell=\frac25}
    \gtrsim 1\, \text{K} .
\end{equation}

The {\it compressible} ES at $\nu = \frac{11}{15}$, or equivalently 
the system of two layers each at
$\frac{11}{30}$ filling, spontaneously ``phase-separates'' into two 
{\it incompressible} FQH liquids at unequal layer fillings of 
$\frac13$ and $\frac25$.  The symmetry-breaking interlayer charge
transfer that must occur [see Fig.\ 3(insets)]
to produce the required density imbalance
entails a significant electrostatic penalty 
${\cal E}_{\text{CAP}}$, but proceeds
only because this capacitive energy barrier is surmounted by the
correlation energy advantage ${\cal E}_{\text{COR}}$
in forming two stable FQH liquids. 
That our best estimates indicate the competing energies
${\cal E}_{\text{CAP}}$ and ${\cal E}_{\text{COR}}$
are comparable attests to the quantitative
plausibility of this explanation.

\begin{figure}[t]
\centerline{\epsfxsize=6.3cm \epsfbox{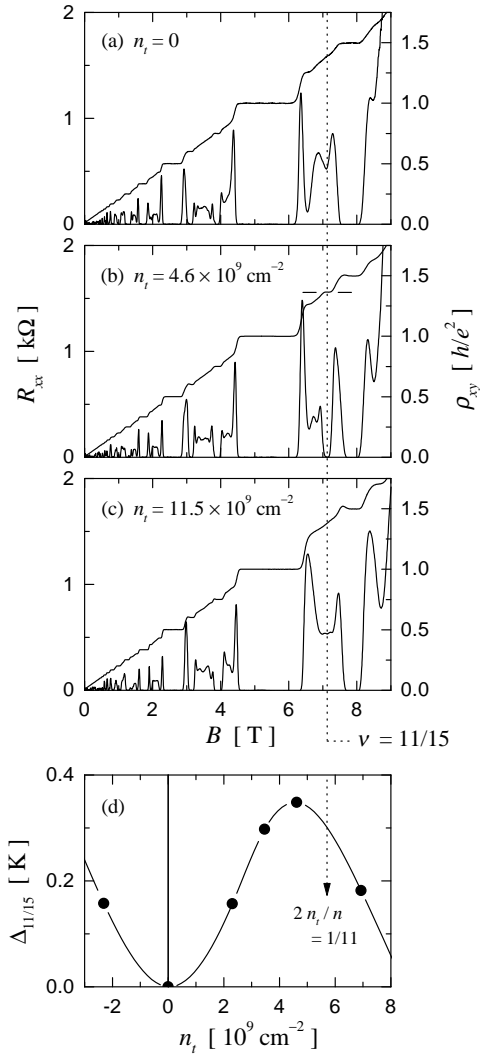}}
\vspace{0.15cm}
\caption{Intentionally-imposed interlayer charge transfer
$n_t$ will stabilize (b)
an incompressible layered $\left(\frac13 + \frac25\right)$ FQH state at
$\nu = \frac{11}{15}$ when the charge distribution is imbalanced close
to the expected $(5:6)$ layer density ratio [dotted line in (d)].}
\label{fig4}
\vspace{-0.38cm}
\end{figure}

Finally, to verify our conjecture that the $\nu = \frac{11}{15}$
state is indeed stabilized by interlayer charge transfer, we did the
following test at $\theta = 0$.  Suppose we start with the ES at an
$n$ where the $\nu = \frac23$ FQH state has just become
2C so that the incompressible state at $\frac{11}{15}$ has barely
developed [e.g., $n = 12.6 \times 10^{10}$ cm$^{-2}$; see
Fig.\ 4(a)].  Now suppose we keep $n$ fixed but
intentionally {\it impose} an interlayer charge transfer
$n_t$ by applying a perpendicular {\it electric} field
(physically generated via front- and back-gate biases of opposite
sign).  As we transfer charge, the $\frac{11}{15}$ FQH state should
get stronger as $2n_t/n$ approaches the ratio
$\left(\frac25 - \frac13\right) / \left(\frac25 + \frac13\right) =
\frac{1}{11}$, and then should become weaker once $2n_t/n$
exceeds $\frac{1}{11}$.  The data shown in Fig.\ 4(a--c) demonstrate that
this behavior is indeed observed in our experiment.  In particular, the
quasiparticle excitation  gap $\Delta_{11/15}$
measured for the $\frac{11}{15}$
FQH state is largest when the $2n_t/n$ ratio is close to
$\frac{1}{11}$, i.e.\ layer densities imbalanced in the ratio
$\left(\frac13 : \frac25 \right) = (5:6)$ [Fig.\ 4(d)].

Two additional features of the data in Figs.\ 1 and 2 are noteworthy.
First, the $\nu = \frac{11}{15}$ state appears to become weaker
with increasing $\theta \gtrsim 40^\circ$.  This is reasonable
and stems from the fact that spontaneous charge transfer will only
occur if 
$\left({\cal E}_{\text{COR}} - {\cal E}_{\text{CAP}}\right) > 0$. At
very large $B_\|$ (or $n$), the two  layers 
become increasingly more isolated and
the capacitive energy opposing charge transfer begins to dominate any 
correlation energy savings
that would come from a $\left(\frac13 + \frac25\right)$ state.
Thus, the system remains compressible, as expected for 
two distant and weakly-coupled parallel 2D ESs at
$\nu = \frac{11}{15}$ (${\frac{11}{30}}$ filling in
each layer).  Second, the $R_{xx}$ minimum near $\nu =
\frac{29}{35}$ $\left(=\frac25 + \frac37\right)$
suggests  a developing FQH state at
this filling [Figs.\ 1(a) and 2]. 
Such a state can be stabilized if, at $\nu =
\frac{29}{35}$, there is an interlayer charge transfer so that one
layer supports a FQH state at $\frac25$ filling and the other at
$\frac37$.  Similarly, the weak $R_{xx}$ minimum observed near
$\nu = \frac{13}{21}$ [Fig.\ 1(a)] may hint at a developing FQH state
stabilized by the formation of $\frac13$ and $\frac27$ FQH states
in the separate layers.

To summarize, we present evidence that a bilayer ES near the magnetic
quantum limit undergoes a
correlation-driven interlayer charge transfer at 
$\nu = \frac{11}{15}$ and $\frac{19}{15}$, and possibly at $\nu =
\frac{29}{35}$ and $\frac{13}{21}$. 
In closing, we note an elegant magnetic analogy for our observations. 
Our bilayer ES can be 
conveniently mapped onto a pseudospin-$\frac12$ system by identifying
the symmetric/antisymmetric subband states (front/back layer states)
with the eigenstates of the $x$-component
($z$-component) of the pseudospin operator 
\cite{Narasimhan:PRB:BiWC}.  Within this framework, a layer imbalance
is equivalent to a non-zero pseudomagnetization $m_z = 2n_t/n$
along the $z$-axis.  
As an increasing $B_z \equiv B_\perp$ 
is applied,
the layers remain balanced ($m_z = 0$)
for all applied fields except at the special
fillings $\nu = p/q$ listed above, when the system
locks to a spontaneous fractional pseudomagnetization 
$m_z = 1/p$ (e.g., $m_z= \frac1{11}$ at $\nu = \frac{11}{15}$,
$m_z =\frac1{19}$ at $\nu = \frac{19}{15}$, etc.) and
behaves as a ``quantized
paramagnet'' with susceptibility 
$\chi_m \equiv n\Phi_{\text{o}} m_z / B_z = 1/q$  
($\Phi_{\text{o}} \equiv h/e$ is the flux quantum).

We thank S.\ M.\ Girvin, S.\ Das Sarma, B.\ I.\ Halperin,
and A.\ H.\ MacDonald for
resplendent conversations.
M.~S.\ acknowledges support by the Alexander von Humboldt 
Foundation.
This work was supported by the NSF.


\begin{references}
\bibitem{Explain1C2C}
More precisely, 1C FQH states are constructed solely from the symmetric
(lower) subband while
in the case of 2C states the antisymmetric subband is also mixed into the
ground state wavefunction.
\bibitem{Tunneling}
Clearly, substantial tunneling would destabilize any $\nu =
\frac{11}{15}$ FQH
state since there is no observable 1C incompressible state at
this filling (see \cite{Monolayer}).
\end{references}
\end{document}